\documentclass{aastex}
\usepackage{emulateapj5}

\newcommand{\etal}{et~al.\ }

\newcommand{\cmsq}{\hbox{cm$^{-2}$}}

\newcommand{\flux}{\hbox{erg~cm$^{-2}$~s$^{-1}$}}
\newcommand{\lumin}{\hbox{erg~s$^{-1}$}}

\newcommand{\aox}{$\alpha_{\rm ox}$}
\newcommand{\nh}{\hbox{${N}_{\rm H}$}}

\newcommand{\chandra}{{\emph{Chandra}}}
\newcommand{\cxo}{{\emph{Chandra X-ray Observatory}}}
\newcommand{\xmm}{\emph{XMM-Newton}}
\newcommand{\asca}{{\emph{ASCA}}}
\newcommand{\rosat}{\emph{ROSAT}}
\newcommand{\hst}{\emph{HST}}

\newcommand{\cross}{Q~2237+0305}
\newcommand{\pg}{PG~1115+080}
\newcommand{\hetwo}{HE~2149$-$2745}
\newcommand{\hezero}{HE~0230$-$2130}
\newcommand{\heone}{HE~1104$-$1805}
\newcommand{\clover}{H~1413+117}
\newcommand{\lbqs}{LBQS~1009$-$0252}
\newcommand{\rxj}{RX~J0911.4+0551}
\newcommand{\hs}{HS~0818+1227}
\newcommand{\qone}{Q~1208+101}
\newcommand{\apm}{APM~08279+5255}
\slugcomment{Accepted by ApJ}
\shorttitle{A Study of QSO Evolution in the X-Ray Band with the Aid of
Gravitational Lensing}
\shortauthors{DAI ET AL.}

\begin{document}

\def\sarc{$^{\prime\prime}\!\!.$}
\def\arcsec{$^{\prime\prime}$}
\def\ls{\lower 2pt \hbox{$\;\scriptscriptstyle \buildrel<\over\sim\;$}} 
\def\gs{\lower 2pt \hbox{$\;\scriptscriptstyle \buildrel>\over\sim\;$}}

\title{A Study of QSO Evolution in the X-ray Band with the Aid of
Gravitational Lensing}

\author{Xinyu Dai, George Chartas, Michael Eracleous, and Gordon P. Garmire}

\affil{Department of Astronomy and Astrophysics,
Pennsylvania State University, University Park, PA 16802}

\email{xdai, chartas, mce, garmire @astro.psu.edu}

\begin{abstract}

We present results from a mini-survey of relatively high redshift
($1.7 < z < 4$) gravitationally lensed radio-quiet quasars observed
with the \cxo\ and with \xmm.  The lensing magnification effect allows
us to search for changes in quasar spectroscopic and flux variability
properties with redshift over three orders of magnitude in
intrinsic X-ray luminosity.  It extends the study of quasar properties
to unlensed X-ray flux levels as low as a few times $10^{-15}$\flux\
in the observed 0.4--8 keV band.  For the first time, these
observations of lensed quasars have provided medium to high
signal-to-noise ratio X-ray spectra of a sample of relatively
high-redshift and low X-ray luminosity quasars.  We find a possible
correlation between the X-ray powerlaw photon index and X-ray
luminosity of the gravitationally lensed radio-quiet quasar sample.
The X-ray spectral slope steepens as the X-ray luminosity increases.
This correlation is still significant when we combine our data with
other samples of radio-quiet quasars with $z > 1.5$, especially in the
low luminosity range between $10^{43}$--$10^{45.5}~\lumin$.  This
result is surprising considering that such a correlation is not found
for quasars with redshifts below 1.5.  We suggest that this
correlation can be understood in the context of the hot-corona model
for X-ray emission from quasar accretion disks, under the hypothesis
that the quasars in our sample accrete very close to their Eddington
limits and the observed luminosity range is set by the range of black
hole masses (this hypothesis is consistent with recent predictions of
semi-analytic models for quasar evolution).  The upper limits of X-ray
variability of our relatively high redshift sample of lensed quasars
are consistent with the known correlation between variability and
luminosity observed in Seyfert 1s when this correlation is
extrapolated to the larger luminosities of our sample.

\end{abstract}

\section{Introduction}
It is important to extend the study of quasars to high redshifts in
order to understand the evolution of quasars and their environments.
One of the main results of recent studies of quasar evolution is that
the quasar luminosity function evolves strongly with
redshift. (e.g. Boyle \etal 1987).  Many studies of quasar evolution
are aimed at explaining this luminosity evolution.  The X-ray band
probes the innermost regions of the central engine of the Active
Galactic Nuclei (AGN).  The study of AGN in X-rays may possibly answer
the question of whether there is an evolution in their central engines
and how this is related to the evolution of the quasar luminosity
function.  The observed X-ray continuum emission of AGN is generally
modeled by a power law of the form $N(E)=N_{0}(E/E_{0})^{-\Gamma}$
where $N(E)$ is the number of photons per unit energy interval.
Extensive studies of AGN during the past decade indicate that this
power-law component is produced by Compton scattering of soft photons
by hot electrons in a corona \citep[e.g.,][]{h93,h94}.  The study of
this power-law component, its correlations with other AGN parameters,
and its evolution reveals important information on the accretion
process of the central object.  The X-ray photon indices of
low-redshift radio-quiet quasars are measured to have mean values of
$\sim$2.6--2.7 in the \rosat\ soft X-ray band \citep{l97,y98} and
$\sim$1.9--2.0 for the \asca\ hard X-ray band \citep{g00,r00}, and are
correlated with the $\rm{H}\beta$ FWHM \citep{r00}.  There is no
strong evidence that the X-ray power-law index evolves with redshift
or correlates with X-ray luminosity to date \citep{g00,r00}.  Another
important parameter that describes the broad band spectral shape of
quasars is the optical-to-X-ray spectral index, quantified as
$\alpha_{\rm ox} =
\log(f_{\rm{2keV}}/f_{2500{\AA}})/\log(\nu_{\rm{2keV}}/\nu_{2500~\AA})$,
where $f_{{\rm 2keV}}$ and $f_{2500~\AA}$ are the flux densities at 2
keV and 2500~\AA\ in the quasar rest-frame, respectively.  Recently,
several studies have provided estimates of \aox\ for very high
redshift ($z > 4$) quasars \citep{vbs03,v03a,v03b,b03}.  
\citet{vbs03,v03a,v03b} found that \aox\ is mainly dependent on the ultra-violet luminosity while \citet{b03} found that \aox\ primarily evolves with redshift.
In
addition to spectral studies, variability studies of Seyfert galaxies
show that the variability amplitude (excess variance) is
anti-correlated with X-ray luminosity \citep{n97,l99}.  Variability
studies have been extended to quasars by several groups
\citep{g00,a00,m02}.  Low redshift quasars ($z < 2$) are found to have
an excess variance consistent with the luminosity relation found in
Seyfert 1s and there is a possible upturn of X-ray variability for
high redshift quasars with $z > 2$ \citep{a00,m02}.

It is also important to compare the properties of quasars near the
peak of their comoving number density, thought to have occurred at
$z\sim2$, with low redshift quasars.  This comparison may provide
clues as to what caused the dramatic decay of the quasar number
density as the Universe expanded.

Most of the observational and analysis techniques employed to date to
study the evolution and emission mechanism of faint, high-redshift
quasars are based on either summing the individual spectra of many
faint X-ray sources taken from a large and complete sample or
obtaining deep X-ray observations of a few quasars.  Although these
techniques may yield important constraints on the average properties
of high redshift quasars they each have significant limitations.

Gravitational lensing provides an additional method for studying
high-redshift quasars.  The extra flux magnification, from a few to
$\sim$100, provided by the lensing effect enables us to obtain
high signal-to-noise ratio (S/N) spectra and light-curves of distant quasars with less
observing time and allows us to search for changes in quasar
spectroscopic properties and X-ray flux variability over three orders
of magnitude in intrinsic X-ray luminosity.  With the aid of lensing,
we can probe lower flux levels than other flux limited samples with
similar instruments and exposures.  This could be an important factor
because the X-ray properties of high-redshift, low-luminosity quasars
could be different from those of other quasars and by studying them we
could possibly obtain information about the evolution of quasars.

Similar lensing studies have also been performed in the sub-millimeter
and CO bands \citep{b02,b02a}, which suggested that gravitational
lensing could be an efficient method for studying various properties
of high-redshift quasars.  Here we present the results of an X-ray
mini-survey of relatively high redshift gravitationally lensed
radio-quiet quasars.  We chose only the radio-quiet quasars in our
sample because the powerful relativistic jets in the radio-loud
quasars will introduce additional complication when modeling the
continuum X-ray emission from the accretion disc.

We use a $H_{0}$ = 50 km s$^{-1}$ Mpc$^{-1}$ and $q_{0} = 0.5$ cosmology throughout the paper.

\section{Observations and Data Reduction}
Our mini-survey contains eleven gravitationally-lensed, radio-quiet
quasars with redshifts ranging between 1.695 and 3.911.  Five of them
contain Broad Absorption Lines (BALs) or mini-BALs in their rest-frame
ultraviolet spectra.  Most of the lensed quasars of our sample were
observed with the Advanced CCD Imaging Spectrometer (ACIS; Garmire
\etal 2003) onboard \chandra\ as part of a Guaranteed Time Observing
program (Principal Investigator: G. Garmire).  The data for two of the
them were obtained through the public \chandra\ archive.  Several of
them were observed twice.  Three of the lensed quasars were also
observed with \xmm.  Table~\ref{tab:olog} presents a log of
observations, including redshifts, Galactic column densities, and
exposure times.  Each observation was performed continuously with no
interruptions.

All of the sources observed with \chandra\ were placed near the aim
point of the ACIS-S array, which is on the back-illuminated S3 chip.
All of the data were taken in FAINT or VERY FAINT mode.  The
\chandra\ data were reduced with the CIAO 2.3 software tools provided
by the \chandra\ X-Ray Center (CXC) following the standard threads on
the CXC website.  Only photons with standard \asca\ grades of 0, 2, 3,
4, 6 were used in the analysis.  We used events in the 0.4--8 keV
energy range in the spectral analysis and events in the 0.2--10 keV
range for the variability studies.  The source events were extracted
from circles with radii ranging from 3\arcsec\ to 5\arcsec\ depending
on the separations of the lensed images of each quasar.  The circles
included all of the lensed images.  Background events were extracted
from annuli with inner and outer radii of 10\arcsec\ and 30\arcsec,
respectively, centered on the sources.  We adjusted the inner and
outer radii of the background subtraction annuli in some cases to
avoid other sources in the field.  The background contributes an
insignificant amount to the count rate in a source region, even during
a background flare.  The average exposure time for the \chandra\
observations was about 25 ks.  The detected source count rates ranged
between 0.003--0.08 $\rm{s^{-1}}$.

Three targets were observed with the European Photon Imaging Camera PN
and MOS detectors \citep{s01,t01} onboard \xmm.  The \xmm\ data were
analyzed with the standard analysis software, SAS 5.3.  The tasks
\verb+epchain+ and \verb+emchain+ from SAS were used to reduce the PN
and MOS data and photons of patterns $\le4$ and $\le12$ were selected
from the PN and MOS data, respectively.  The \xmm\ data are affected
more than the \chandra\ data by background flares because the Point
Spread Function of \xmm\ is significantly larger than that of
\chandra-ACIS.  Several strong background flares occurred during the
\xmm\ observations.  These flares are filtered out in the spetral
analysis.

\section{Spectral Analysis}
\subsection{Power-law Continuum}
We performed spectral fitting in order to obtain the power-law indices
of the X-ray continuum emission components of the quasars in our
sample.  The photon indices that we measured were in the observed
0.4--8 keV energy range.  We used the \chandra\ data for all of the
lensed quasars in the spectral analysis except \pg, where we used the
\xmm\ data.  \rxj\ and \apm\ were also observed with \xmm.  
We did not
use these data, however, because there were large amplitude and long duration
 background flares in
the \xmm\ observations.
The background flare in the \xmm\ observation of \rxj\ almost spans the
entire observation.
The background flare in the \xmm\ observation of \apm\ occurs during the last $\sim20$ ks of the observation.
We filtered the background flare time from this observation
and performed a spectral analysis to compare with our \chandra\ results.
For our later correlation analysis, we
used the \chandra\ data of \apm\ to avoid the possible complication from
the cross calibration between \chandra\ and \xmm\ instruments.
For
those sources observed twice with \chandra, we performed simultaneous
fits to the spectra extracted from the two observations except in the
case of \apm, where we used only the second observation because it was
much longer than the first one.

We extracted spectra for each quasar using the CIAO tool
\verb+psextract+.  We extracted events from all of the lensed images
for each source except for the case of \clover. A possible
microlensing event in \clover\ appears to amplify and distort the
spectrum of image A only and thus affects the spectral slope greatly
\citep{c03}.  In this case, we extracted the spectrum of the
microlensed image A and the spectrum of the other three images
separately.  A microlensing event is also detected in \cross.
However, the spectral shape of the microlensed image A of \cross\ is
not significantly affected by the microlensing event \citep{d03}.

Spectral fitting was performed with \verb+XSPEC V11.2+ \citep{a96}.
There are typically several hundreds (180--6000) detected source
events in the spectra of the target quasars, except for the spectrum
of \hetwo\ which has only 23 detected events.  The moderate S/N of our
spectra allows us to fit each of them individually using relatively
complex models.  Thus we can constrain the underlying power-law slopes
more accurately than in previous studies of unlensed quasars of
similar redshift.  The models we used are listed in Column~3 of
Table~\ref{tab:spec}.  All spectral models included an underlying
power-law model modified by Galactic absorption.  The Galactic
absorbing columns were obtained from \cite{d90}.

\begin{figure}
\epsscale{1}
\plotone{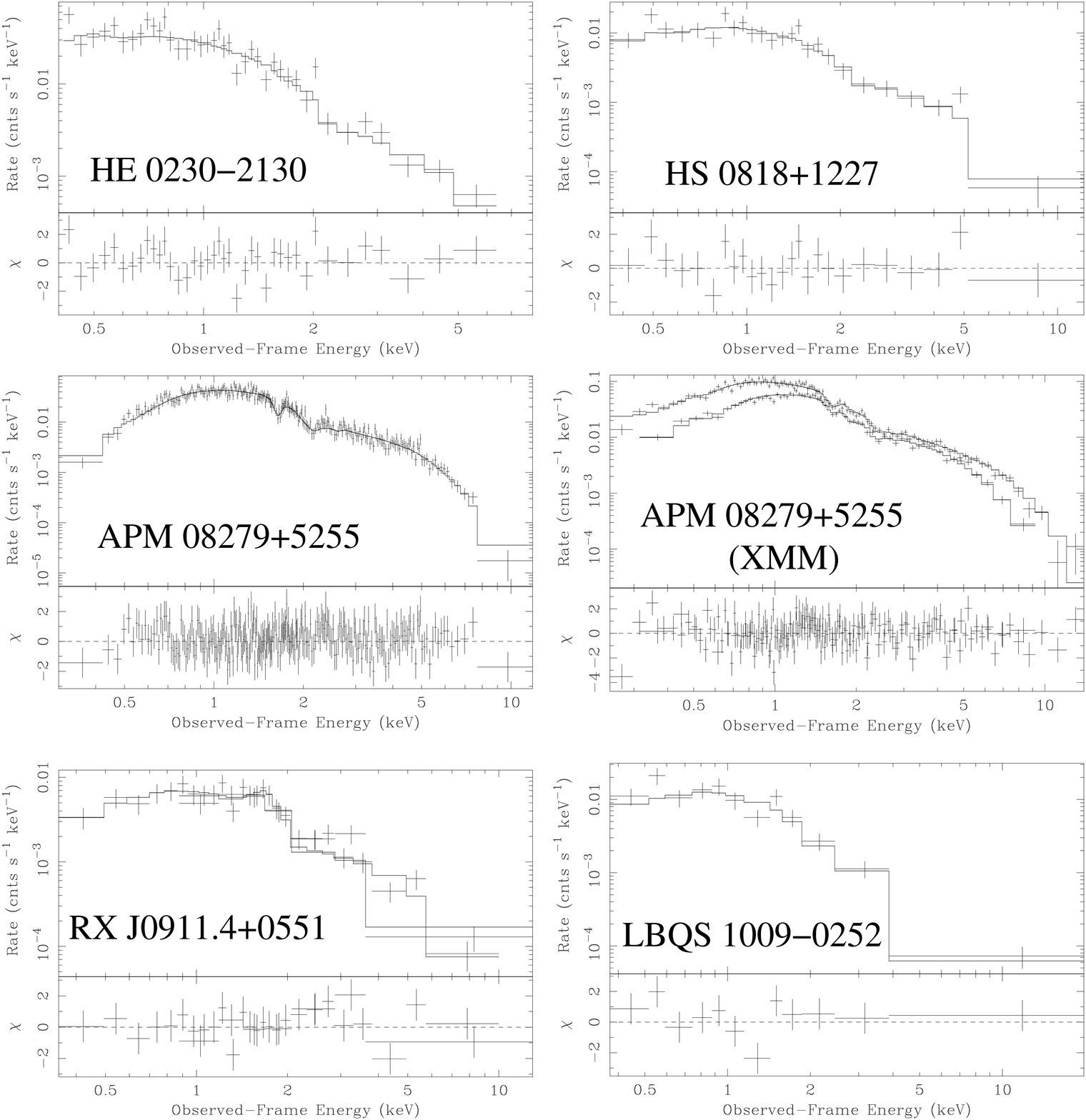}
\caption{Spectra of gravitationally lensed radio-quiet quasars observed with \chandra, except for the spectra of \apm\ and \pg\ that were observed with \xmm. (Part I)\label{fig:spec}}
\end{figure}

To account for the recently observed quantum efficiency decay of ACIS, 
possibly caused by molecular contamination of the ACIS filters, 
we have applied a time-dependent correction to the
ACIS quantum efficiency implemented in the \verb+XSPEC+ model \verb+ACISABS1.1+.\footnote{ACISABS is an XSPEC model contributed to the \chandra\ users
software exchange web-site
http://asc.harvard.edu/cgi-gen/cont-soft/soft-list.cgi.}  If the
source was a BALQSO with a medium S/N spectrum, we added a neutral
absorption component at the redshift of the source.  For high S/N
BALQSO spectra such as \pg\ and \apm\ we used the \verb+absori+ model
in \verb+XSPEC+ to model the intrinsic absorption component as an ionized absorber
\citep{c02,cbg03}.  We added a neutral absorption component at the
redshift of the lens for \cross\ \citep{d03}.  For some of the spectra
containing emission or absorption line features, we added Gaussian
line components to model them accordingly.  The spectra of the quasars
are displayed in Figures~\ref{fig:spec}.  The spectral fitting results
are given in Table~\ref{tab:spec}.  In Table~\ref{tab:spec}, we also
list the observed 0.4--8 keV fluxes and rest-frame 0.2--2 keV and 2-10
keV luminosities for the lensed quasars in the sample not corrected
for the lensing magnification.  The unlensed luminosities are
calculated in $\S$\ref{sec:mag}.  We have corrected for the various 
absorption components when calculating X-ray luminosities.  
These absorption components include Galactic absorption, absorption 
in the lensing galaxy, intrinsic absorption, and the ACIS contamination.

The mean photon index of the radio-quiet quasars in our sample is 
 $1.78\pm0.06$
with a dispersion of $0.27\pm0.08$ and the median photon index of the
radio-quiet quasars in our sample is 1.86.

\subsection{\aox}
We calculated the optical-to-X-ray power-law slope, \aox, for the
quasars in our sample.  The differential magnification between the
optical and X-ray band is insignificant because both of the source
regions are estimated to be much smaller than the Einstein radius.  The rest-frame 2
keV flux densities were calculated from the best-fit models to the
spectra of the quasars.  The redshifts of our sample range from 1.7 to
4, thus the rest-frame 2 keV energy falls in the 0.4--0.74 keV
observed-frame energy range.  We removed all of the absorption
components including the intrinsic ones for the BALQSOs when
calculating the rest-frame 2 keV flux densities in order to study the
properties of the intrinsic continuum.  The rest-frame 2500~\AA\
fluxes were obtained from the published optical data in the literature
and from current ongoing programs such as the CfA-Arizona Space
Telescope LEns Survey (CASTLES)
\footnote{The CASTLES website is located at http://cfa-www.harvard.edu/glensdata/.}
 and the Optical Gravitational Lensing Experiment (OGLE).
\footnote{The OGLE website is located at
http://bulge.princeton.edu/\~{}ogle.}  We first converted the optical
magnitudes to flux densities at the effective wavelengths of the
filters, then extrapolated them to the rest-frame 2500~\AA\ flux
densities.  We used the standard relations between magnitudes and flux
densities for the V and R bands.  For the F814W magnitudes obtained
with the \hst\ WFPC2, we used the relation from \cite{h95} to convert
between magnitudes and flux densities.  We used
$f_{\nu}\propto\nu^{-0.7}$ in the extrapolation.  Typical quasar
optical spectral indices are in the -0.5 to -0.9 range \citep{sch01}.
We chose the optical magnitudes from the bands closest to the
redshifted 2500~\AA\ wavelength to reduce the error in the
extrapolation.  We corrected the Galactic extinction based on
\cite{sch98}.  The extinction induced by the dust from the lensing
galaxies is not significant except for \cross, which is lensed by the
central bulge of the galaxy.  We corrected it according to the
appendix of \citet{ag00}.  The \aox\ values are listed in
Table~\ref{tab:aox}.  The listed error-bars of \aox\ are at the
$1\sigma$ level and account for the uncertainties in the X-ray
spectral slope and normalization, the conversion from optical
magnitude to flux density and the extrapolation of the optical/UV
spectrum.  We note that a source may have varied between the optical
and X-ray observations thus leading to systematic errors in our
estimates of \aox.  For the range of \aox\ in our sample, a change of
optical flux by a factor of two will lead to a change of
$\Delta$\aox$\sim0.1$.  The rest-frame 2~keV and 2500\AA\ flux
densities and rest-frame 2500\AA\ luminosity densities are also listed
in Table~\ref{tab:aox}.  These values are corrected for lensing
magnification as we discuss further in $\S$\ref{sec:mag}.  In
conclusion we find that the mean \aox\ value for all the quasars in
our sample is $-1.70\pm0.02$ ($-1.66\pm0.02$ without \hetwo\ which has only 23 X-ray events) with a dispersion of $0.20\pm0.03$.

\section{Magnification and Intrinsic X-ray Luminosities}
\label{sec:mag}

We estimated the magnification for each of the lensed systems in order
to obtain the unlensed X-ray luminosities of the quasars.  We first
searched in the literature for magnifications of well-modeled systems.
However, not all of the systems in our sample have been previously
modeled in detail and for some cases the magnification values are not
included in the published analysis.  We modeled these systems with the
\verb+gravlens+ 1.04 software tool developed by C. Keeton
\citep{k01a,k01b}.  We used a singular isothermal elliptical (SIE)
mass profile with external shear in most cases, and took into account
optical image positions and flux ratios obtained from the CASTLES
survey.  We assumed a 20\% error-bar for the flux ratios to account
for uncertainties from microlensing or differences between optical and
X-ray flux ratios.  Our modeling results are listed in
Table~\ref{tab:mag}.  We compared the magnification values that we
obtained with those published in \cite{b02} and found them to be in
good agreement.  We note that the magnification values obtained may
have large systematic errors because the SIE model adopted in this
analysis may not be suitable for all the systems.  For example, in
\pg, a set of different mass potentials are used in the modeling of
the magnification in \cite{i98}.  A magnification range of 20--46 was
obtained for this system.  Based on this example, we adopt a factor of
two as a typical systematic uncertainty in the magnification.  In
$\S$\ref{sec:disli} we investigate how this systematic uncertainty
could affect for our conclusions via Monte Carlo simulations.

\section{Variability Analysis}
The light-curves for the \chandra\ observations of the lensed quasars
of our sample are displayed in Figure~\ref{fig:lc} binned with a bin
size of 1000 s (observed-frame).  We show the light-curves of the
longer \chandra\ exposure if two \chandra\ observations of the same
target are available.  We excluded \hetwo\ in our timing analysis
because of its low S/N (only 23 events).

To estimate the relative variability of the light-curves, we used the normalized excess variance \citep{n97,t99,e02} defined as
\begin{equation}
\sigma^{2}_{\rm rms} = \frac{1}{N\mu^{2}}\sum_{i=1}^N [(X_{i}-\mu)^{2}-\sigma_{i}^{2}]
\end{equation}
where $N$ is the number of bins in the light-curve, $X_{i}$ are the
count rates per bin with error $\sigma_{i}$, and $\mu$ is the mean
count rate.  The error on $\sigma^{2}_{\rm rms}$ is given by
$S_{D}/[\mu^2(N)^{1/2}]$, where
\begin{equation}
S_{D}^{2} =
\frac{1}{N-1}\sum_{i=1}^{N}\{[(X_{i}-\mu)^{2}-\sigma_{i}^{2}]-\sigma^{2}_{\rm
rms}\mu^{2}\}^{2}
\end{equation}
This method normalizes the variability amplitude to the flux, thus is
less biased towards the low flux light-curves.  It is better to apply
this method to a sample of light-curves of similar lengths to avoid
possible bias.  For example, it would be easier to detect long
time-scale variability from a light-curve with a long duration.  In
addition, the bin sizes should be similar in order to compare
variability on similar time-scales.  High S/N light-curves are needed
for this analysis as the variability could be easily dominated by
noise in the low S/N light-curves.  This excess variance method also
has some limitations.  One of them is that it is not very sensitive to
short flares of moderate amplitude.  The excess variance values for
\chandra\ light-curves with three different bin sizes are listed in
Table~\ref{tab:evar}.  We also calculated the excess variance for the
three sources observed with \xmm\ and the results are listed in
Table~\ref{tab:evar2}.  The background was subtracted in the
calculation.  Overall, the error-bars on the excess variance are quite
large and in some cases, the values are negative.  Therefore, for most
quasars we can really only obtain an upper limit on the value of the
excess variance.

\section{Results and Discussion}
\subsection{Luminosity and Spectral Index}
\label{sec:disli}
The unlensed 0.2--2 keV and 2--10 keV X-ray luminosities of the lensed
quasars in our mini-survey range from $10^{43}$ to $10^{46}$~\lumin.
The mean spectral index of our sample, 1.78, is consistent with the
value of 1.84 from the recent study of very high redshift quasars
\citep{v03b} and the value of 1.89 from the radio-quiet \asca\ sample
\citep{r00}, especially when one considers the large dispersion of
spectral indices in these studies and our small sample size.

We plot the rest-frame 0.2--2 keV and 2--10 keV X-ray luminosity
against redshift in Figure~\ref{fig:gamma}(a), and the photon indices
vs. redshift, 0.2--2 keV luminosity, and 2--10 keV luminosity in
Figures~\ref{fig:gamma}(b), (c), and (d), respectively.  Figures
~\ref{fig:gamma}(c) and (d) show a correlation between the spectral
indices of our sample of lensed quasars and their X-ray luminosities.
We tested the significance of this correlation using the Spearman's rank
correlation method.  We obtained a rank correlation coefficient of
0.94 significant at the greater than 99.997\% confidence level for the
correlation between photon index and 0.2--2 keV luminosity and a
coefficient of 0.71 significant at the 98.6\% confidence level for the
correlation between photon index and 2--10 keV luminosity.  
We note that the measured rest-frame 
0.2--2 keV luminosity for high redshift quasars obtained from 
\chandra\ and \xmm\ data is less accurate than the rest-frame 2--10 keV
luminosity because it depends on extrapolation and suffers from the
uncertainty in the absorbing columns towards the quasars.  
We adopt a significance at the 98.6\% confidence level 
for this correlation, 
evaluated with the 2-10 keV luminosities.  We also
tested for a possible correlation between X-ray luminosity and
redshift and between photon index and redshift and found none.  

The correlation between the photon index and X-ray luminosity was a
surprising result.  Therefore, we tested if this correlation is driven
by certain data points.  Excluding one of our data points at a time
and performing the Spearman's rank correlation between $\Gamma$ and
the 0.2--2 keV luminosity to the rest of the data set, we find a
strong correlation with a significance level greater than 99.98\% each
time.  We performed a similar analysis with the 2--10 keV luminosity,
and the correlation is significant at the greater than 95\% confidence 
level each time.
This indicates that the correlation is not driven by any
particular data point.  
We further performed correlation tests by
excluding data points with errors in $\Gamma$ larger than 0.20.  Two data 
points are excluded and the rest of the data points still show 
a correlation 
significant at the greater than 99.98\% confidence level.  
The significance is 
at the 92\% confidence level when 2--10 keV luminosities are used.
We also tested if
there is a bias from different flux levels in our sample such that
quasars with low fluxes have a systematically flatter slope because of
measurement errors.  In principle, this should not be a problem for
the sample in this paper because some low luminosity quasars have high
S/N spectra as a result of the gravitational lensing effect.  To test
this, we simulated spectra with the same photon index ($\Gamma = 2$)
but with a wide range of S/N and then fitted them with the same models
within \verb+XSPEC+.  We found that for low S/N spectra, when the
total number of photons is less than $\sim$200, the measured $\Gamma$
could possibly be flatter by $\sim0.1$.  This would affect three of our
data points: \hetwo, \lbqs, and \qone\ with (one
low-luminosity and two high-luminosity quasars).  We tested this effect
by increasing the photon index by 0.1 for these three quasars and performed
the correlation test again and found
the correlations between $\Gamma$ and
0.2--2 keV (2--10 keV) luminosities are significant at the 99.99\% (98.7\%)
confidence levels, respectively.

A $\Gamma$--${L_{X}}$ correlation has been previously searched for
in other samples of radio-quiet quasars.  \citet{r97} reported first a
possible correlation between $\Gamma$ and luminosity with nine
radio-quiet quasars observed with \asca.  However, this correlation
was later not found in the study of a larger sample of 27 radio-quiet
quasars observed with \asca\ (including the original nine) by
\citet{r00}.  \citet{g00} also searched for a correlation between
$\Gamma$ and luminosity in another sample of radio-quiet quasars
observed with \asca\ and did not find one.  However, the quasars in
our sample all have relatively high redshifts $1.7 < z < 4$ compared
to the redshifts of quasars incorporated in previous studies of the
$\Gamma$--$L_{X}$ correlation.  There is only one out of 26
quasars with $z > 1.5$ in the sample of \citet{g00}, and there are six
out of 27 quasars with $z > 1.5$ in the sample of \citet{r00} and most
of them have large errors on $\Gamma$.  In addition, the high redshift
quasars in the sample of \citet{g00} and \citet{r00} have high
luminosities and span a small luminosity range of
$10^{45}$--$10^{46}~\lumin$.  The sample of lensed quasars in this
paper have lower luminosities and span a larger luminosity range of
$10^{43}$--$10^{45}~\lumin$.  Recently, \citet{p03} presented another
sample of quasars observed with \xmm\ which contains several high
redshift, low-$L_{X}$ quasars.  We plotted the
$\Gamma$--$L_{X}$ diagram for quasars of $z > 1.5$ from the
sample of \citet{p03}, \citet{r00}, \citet{g00}, and \citet{v99} in
Figure~\ref{fig:gamma1}.  We excluded one data point from \citet{p03}
with a large measurement error on $\Gamma$.  We also excluded PHL 5200
from \citet{r00} in our analysis because new \xmm\ observations of the
object showed that most of the X-rays originate from a nearby radio
source \citep{bfg02}.  We corrected the X-ray luminosity of \heone\ reported in \citet{r00} to account for the gravitational lensing magnification of about 12 based on our analysis presented in 
$\S$\ref{sec:mag}.  We used rest-frame 2-10 keV luminosities in
order to be consistent with the previous analysis.  Most of the data
from \citet{p03} are consistent with the $\Gamma$--$L_{X}$
correlation found in our sample of lensed quasars, especially in the
low $L_{X}$ range between $10^{43.5}$--$10^{45.5}~\lumin$.  On
the high luminosity end ($L_{X} > 3\times10^{45}~\lumin$) of high
redshift quasars ($z > 1.5$), the $\Gamma$ dependence on $L_{X}$
seems to flatten out or even has an anti-correlation pattern.  We
performed the Spearman's rank correlation test to all of the ($z > 1.5$)
quasars and did not find a correlation between $\Gamma$ and
2--10 keV luminosity.  We separated the quasars in two groups with
luminosities below and above $3\times10^{45}~\lumin$ and performed the
rank correlation again to each group.  The low luminosity quasars show
a strong correlation between $\Gamma$ and 2--10 keV luminosity 
at the 99.97\%
confidence level and the high luminosity quasars show an anti-correlation
between $\Gamma$ and 2--10 keV luminosity at the 98.9\% confidence level.

We performed a Monte-Carlo simulation to test how much the
$\Gamma$--$L_{X}$ correlation is affected by the uncertainty of
the magnification factors of our lensed quasars.  We used the combined
data sets from the lensed sample and from the quasars of \citet{p03}
with $z > 1.5$, which show a strong $\Gamma$--$L_{X}$
correlation.  We assumed a systematic uncertainty on the magnification
of a factor of two (see discussion in $\S$\ref{sec:mag}) and simulated
10,000 data sets with $L_{X}$ perturbed from its measured value
randomly, within the range of this error-bar.  We fixed the data
points from \citet{p03} at their original values.  We performed
the Spearman's correlation test between $\Gamma$ and 2--10 
keV luminosity
on the simulated data sets and found that
the $\Gamma$--$L_{X}$ correlation in each simulated data set is
at least significant at the 98.6\% confidence level for 10,000 simulations
and 9,888 of them have $\Gamma$--$L_{X}$ correlations significant
at the greater than 99.5\% confidence level.  Therefore, it is
unlikely that this correlation is driven by the uncertainties in the
magnification factors.

Possible errors in magnification factors, the small number of quasars in the
present sample combined with the medium S/N of several of the
observations may have led to an unaccounted systematic effect.
Additional observations of $z\sim2$ lensed quasars with better S/N and
a more detailed modeling of all the lenses in our sample will help confirm
this result.  Although the $\Gamma$--$L_{X}$ correlation found in
this paper could be a result of selection effect, we discuss possible
interpretations of this correlation below.

Previous studies did not find any correlation between $\Gamma$ and
$L_{X}$ in low-redshift quasars and Seyfert 1s.  Recently,
several X-ray monitoring programs of Seyfert galaxies have indicated a
correlation between X-ray spectral index and 2--10 keV X-ray flux
\citep{ch00,p00,va01}.  The spectra of these Seyfert galaxies steepen
as the X-ray flux increases.  \cite{va01} suggested that variations of
several properties of the X-ray emitting corona (e.g., electron
temperature, optical depth, size) with the X-ray flux may explain this
$\Gamma$ vs. X-ray-flux relation.  Similarly, a $\Gamma$ vs. X-ray-flux
relation also exists in some Seyfert 2 and radio-loud objects
\citep{g01,z01,gse03}.  These observations indicate that although the
$\Gamma$--$L_{X}$ relation does not apply to a sample of
low-redshift quasars and Seyferts, this relation applies to individual
objects.  In contrast to the low-redshift quasars, the
$\Gamma$--$L_{X}$ dependence found at high redshift, especially
in high-redshift and low-$L_{X}$ quasars indicates that these
quasars may represent a distinct sub-population of quasars.  It is
possible that the properties of the X-ray emitting coronae of the
high-redshift quasars are more homogeneous than those in the
low-redshift quasars.

\subsection{Possible Interpretations of the Correlation Between 
the X-Ray Luminosity and Spectral Index}

We explore two possible interpretations of the correlation between the
spectral index and the X-ray luminosity in the context of the model of
\citet{h93} and \citet{h94}. Our first hypothesis is that there is a
narrow range (of order a few) of accreting black hole masses at high
redshift while the observed range of observed X-ray luminosities
(spanning aproximately two orders of magnitude) is caused by a large
range in the accretion rate, i.e., a wide range of Eddington ratios
(the ratio of the accretion rate to the Eddington rate). Our second
hypothesis is that the opposite is true, namely that the range of
accreting black hole masses at high redshift is large (spanning
approximately two orders of magnitude), while the Eddington rate is
fairly constant and close to unity (i.e., the accretion rate is always
fairly close to the Eddington limit, regardless of the mass).

In the model of \citet{h93} and \citet{h94} the inner accretion disk
is sandwiched by a hot, tenuous, possibly patchy corona which emits
the hard X-ray photons.  The corona cools by inverse Compton
scattering of soft photons ($E\ls 100$~eV) from the underlying disk,
which results in the hard X-ray photons that we observe. \citet{h97}
explored how the model predictions change for a wide range of the
system parameters, namely the optical depth and temperature of the
corona and the temperature of the (assumed) black-body spectrum of
seed photons. In particular, \citet{h97} found that the 2--10~keV
spectral index is nearly proportional to the optical depth in the
hot corona and also decreases as the temperature of seed soft-photon
spectrum increases. Moreover, in the case of a ``compact'' corona,
i.e., one in which pair production is high enough that the Compton
scattering optical depth is dominated by electron-positron pairs, the
2--10~keV spectral index is nearly proportional to the log of the
2--10~keV luminosity, just as we have found observationally.

In our first hypothesis the observed range of luminosities is set by
the range of accretion rates, with the black hole mass being nearly
constant. In this context the observed correlation might be explained
if the optical depth in the hot corona is somehow related to the
accretion rate; for example, the density of the corona could increase
with accretion rate and so would the optical depth.  There is a
competing effect, however: as the accretion rate decreases so does the
temperature of the underlying thin disk, which provides the seed
photons, with the result that the spectral index of the emerging X-ray
spectrum increases.

In our second hypothesis, the accretion rate for all objects is very
close to the Eddington limit, while the observed range in luminosity
is set by a range in black hole masses. In the context of this
hypothesis, and under the assumption that the optical depth of the
corona is dominated by electron-positron pairs, \citet{h97} predict a
correlation between the 2--10~keV spectral index and luminosity just
like the one we observe. The optical depth in this case depends only
on the ``compactness'' of the corona, which can be expressed as $
\ell_c\sim 10^4 (h/r^2) (L_X/L_{Edd})$ \citep{h93}, where
$L_X/L_{Edd}$ is the ratio of the X-ray luminosity to the Eddington
luminosity (roughly proportional to the Eddington ratio) and $h$ and
$r$ are the vertical and radial extent of the corona in units of the
Schwartzschild radius. Thus, in this scenario, the spectral index
could increase if the compactness of the corona increases with black
hole mass (See equation 16 of \citet{h93}), 
with the consequence that the emerging 2-10~keV luminosity
increases as well. 
Moreover, as the optical depth reaches unity, the
spectral index reaches its maximum value and begins to decline for
larger optical depths. This feature may explain the turnover that we
observe in the spectral index $vs$ luminosity plot of
Figure~\ref{fig:gamma1}. We therefore favor this interpretation over
the previous one.  This interpretation is also appealing because it is
consistent with the predictions of semi-analytic models of
\citet{kh00} for the cosmological evolution of supermassive black
holes and their fueling rates. In particular, these authors predict
that at redshifts between 1.5 and 3, the fueling rate of supermassive
black holes in quasars are within a factor of a few of the Eddington
limit, while their masses span the range between $10^7$ and
$10^9$~M$_{\odot}$.

\subsection{The Optical-to-X-Ray Index, \aox}

\citet{vbs03,v03a,v03b} found that the optical-to-X-ray index, \aox, of radio-quiet quasars is mainly dependent on the rest-frame ultra-violet luminosity such that quasars with higher ultra-violet luminosities have steeper \aox\ values (see Figure 5).  They also do not detect a significant redshift dependence of \aox\ in their analysis, however, \citet{b03} found that \aox\ depends primarily on redshift (\aox\ steepens with increasing redshift) and weakly on luminosity.
We compare the average \aox, -1.70 (-1.66 without \hetwo), of our
sample with the average \aox, -1.71, of \cite{v03b}, and the \aox\
values are consistent within the observed r.m.s. dispersion.  The
redshift range of our quasars is $1.7 < z < 4$, lower than the
sample of $ z > 4$ quasars studied in \citet{v03b}.  
However, the consistency between the \aox\ values
from the lensed sample in this paper and the sample of $ z > 4$ quasars studied in \citet{v03b}
does not rule out a possible small evolution of \aox\ with redshift due to our small sample size.
We plotted the \aox\ values from our lensed sample against their redshift 
and rest-frame
2500\AA\ luminosity density in Figure~\ref{fig:aox}(a) and (b),
respectively.  We performed the Spearman's rank correlation test between
\aox\ and $z$ and between \aox\ and $L_{2500~\AA}$, but found no
significant correlation in either case.  Considering the large
dispersion of the \aox\ values from the best fit \aox--$L_{2500\AA}$
relation of \citet{vbs03}, it is not surprising that no significant
correlation is found in our data set because of the small sample size.
We over-plotted our \aox\ data on the \aox--$L_{2500\AA}$ relation of
\citet{vbs03} with a dispersion of 0.25 as indicated by the shaded
region in Figure~\ref{fig:aox}(b), and found that most of our data are
consistent with the \aox--$L_{2500\AA}$ relation except one BAL QSO
data point.

We also searched for possible correlations between the \aox\ values and other
properties of the quasars such as $\Gamma$ and $L_{X}$.  The
correlation results were strongly biased by the -2.1 \aox\ value of
\hetwo.  When we removed this data point, no significant correlation
was detected.

\subsection{Short Time Scale Variability}
We used the upper limits of the excess variance obtained from
light-curves of bin size 1000 seconds when comparing with results for
local Seyferts and other high redshift quasars.  This bin size
corresponds to rest-frame time scales ranging from 200 to 370 s for the quasars
in our sample and is similar to the time scales used in other short
time scale variability studies in Seyfert galaxies and quasars
(e.g. Turner \etal 1999).  We plotted the excess variances of the high
redshift lensed quasar sample together with the variances for Seyfert
1s \citep{n97}, NLS1s \citep{l99}, LLAGN \citep{p98}, and high
redshift ($z > 2$) quasars \citep{m02} in Figure~\ref{fig:excess}.  We
fitted the Seyfert 1 points with a power law and fixed the power-law
index to the same value reported in \citet{n97}, allowing only the
normalization parameter to be free.  The power-law relation is shown
as a solid line in Figure~\ref{fig:excess}.  All of the upper limits
of the excess variances for the high redshift lensed quasars of this
paper are consistent with the Seyfert 1
$\sigma^{2}_{rms}$--$L_{X}$ correlation.  When we compare the
upper limits with the NLS1s data, three of the upper limits of our
sample are located below the $\sigma^{2}_{rms}$--$L_{X}$ relation
for NLS1s, which have about an order of magnitude larger excess
variance than the Seyfert 1s.  The consistency between the excess
variance upper limits of our lensed sample and the Seyfert 1s
$\sigma^{2}_{rms}$--$L_{X}$ relation does not contradict the
possible upturn in the value of the variability of $z > 2$ quasars for
quasars with luminosities of about $10^{46}\lumin$ as presented by
\cite{m02}.  Most of the quasars in our sample have redshifts greater
than 2, but the luminosity range of our sample is below the luminosity
range of \citet{m02}.

The excess variance analysis that we discuss in this paper probes the
shortest time-scale variability of radio-quiet quasars.  There are two
main interpretations that have been presented in previous studies to
explain the excess variance and luminosity relation. (e.g., Nandra
\etal 1997; Manners, Almaini, \& Lawrence 2002).  First, the more
luminous sources may be physically large and lack the short time-scale
variability.  Second, the more luminous sources may contain more
independently flaring regions and these flares contribute less to the
total flux for the more luminous sources compared with the low
luminous sources.  There are several rapid flares observed in our
light-curve sample, in particular, such flares were detected in
\rxj, \cross, and \pg\ \citep{c01,d03,d04}.  
We performed the Kolmogorov-Smirnov test 
to the light-curves of the individual images of the systems, \rxj, \cross, and \pg, and found the light-curves are variable at the 99.8\%,
97\%, and 99.6\% confidence levels, respectively.
These observations support
the second explanation where there are also flares in the high
luminosity quasars, however, due to the quasar's high luminosity,
flares contribute less to the total luminosity and thus luminous
quasars have smaller excess variances.

\section{Conclusions}
We presented results from a mini-survey of relatively high redshift
($1.7 < z < 4$) gravitationally lensed radio-quiet quasars observed
with the \cxo\ and \xmm.  We demonstrated how gravitational lensing
can be used to study high-redshift quasars.  Our main conclusions are
as follows:

\begin{enumerate}
\item {We find a possible correlation between the spectral slope and
X-ray luminosity of the gravitationally lensed quasar sample.  The
X-ray spectral slope steepens as the X-ray luminosity increases.  The
limited number of quasars in the present sample combined with the
medium S/N of several of the observations and the systematic
uncertainties from the lensing magnification modeling may have led to
an unaccounted systematic effect.  Additional observations of $z\sim2$
lensed quasars with better S/N and more detailed modeling of all the
lenses in our sample will allow us to confirm this result.

Such a correlation is not observed in nearby $z < 0.1$ quasars
suggesting that quasars at redshifts near the peak of their number
density may have different accretion properties than low redshift
quasars.  When we combined the data from other samples of radio-quiet
quasars selecting quasars with redshift ($z > 1.5$) together with the
present lensed sample, the correlation is still significant,
especially in the low X-ray luminosity range between
$10^{43-45.5}~\lumin$.  If this correlation is confirmed by future
studies, it could provide significant information on the emission
mechanism and evolution of quasars.}

We suggest that this correlation can be understood if we hypothesize
that the quasars in our sample are fueled at rates near their Eddington
limit (consistent with recent models for quasar evolution) and that
the optical depth of their X-ray emitting coronae increases with black
hole mass.

\item{We did not find a strong correlation between the
optical-to-X-ray spectral index, \aox, on either redshift or on UV
luminosity in our small sample.  However, most of our data points are
consistent with the \aox--$L_{2500\AA}$ correlation of \citet{vbs03}}.

\item{Our estimated upper limits of X-ray variability of the
relatively high redshift lensed quasars sample is consistent with the
known correlation observed in Seyfert 1s.}

\end{enumerate}

\acknowledgements We thank Steinn Sigurdsson and the anonymous referee
for very helpful comments and 
discussions.  We acknowledge the financial support by NASA grant NAS
8-01128.

\clearpage
\begin{deluxetable}{llcccccccc}
\tabletypesize{\scriptsize}
\tablecolumns{7}
\tablewidth{0pt}
\rotate
\tablecaption{The Sample of Gravitational Lensed Radio-Quiet Quasars \label{tab:olog}}

\tablehead{
\colhead{} &
\colhead{} &
\colhead{} &
\colhead{} &
\multicolumn{2}{c}{First \chandra\ Observation} &
\multicolumn{2}{c}{Second \chandra\ Observation} &
\multicolumn{2}{c}{\xmm\ Observation} 
\\
\colhead{} &
\colhead{} & 
\colhead{} & 
\colhead{Galactic \nh\tablenotemark{a}} &
\colhead{Date} &
\colhead{Exposure} &
\colhead{Date} &
\colhead{Exposure} &
\colhead{Date} &
\colhead{Exposure}
\\
\colhead{Quasars} &
\colhead{Redshift} &
\colhead{BAL?} &
\colhead{($10^{20}~\cmsq$)} &
\colhead{} &
\colhead{(ks)} &
\colhead{} &
\colhead{(ks)} &
\colhead{} &
\colhead{(ks)}
}

\startdata
\hezero & 2.162 & \nodata & 2.3 & 2000-10-14 & 15 & \nodata & \nodata & \nodata & \nodata \\
\hs & 3.115 & \nodata & 3.4 & 2002-12-18 & 20 & \nodata & \nodata & \nodata & \nodata \\
\apm & 3.911 & BAL & 3.9 & 2000-10-11 & 10 & 2002-02-24 & 90 & 2002-04-28 & 100 \\
\rxj & 2.80  & mini-BAL & 3.7 & 1999-11-03 & 30 & 2000-10-29 & 10 & 2001-11-02 & 16 \\
\lbqs & 2.74 & \nodata & 3.8 & 2003-01-01 & 10 & \nodata & \nodata & \nodata & \nodata \\
\heone & 2.303 & \nodata & 4.6 & 2000-06-10 & 49 & \nodata & \nodata & \nodata & \nodata \\
\pg & 1.72 & mini-BAL & 3.5 & 2000-06-02 & 26 & 2000-11-03 & 10 & 2001-11-25 & 60 \\
\qone & 3.80 & \nodata & 1.7 & 2003-03-02 & 10 & \nodata & \nodata & \nodata & \nodata \\
\clover  & 2.55  & BAL & 1.8 & 2000-04-19 & 40 & \nodata & \nodata & \nodata & \nodata \\
\hetwo & 2.033 & BAL & 2.3 & 2000-11-18 & 10 & \nodata & \nodata & \nodata & \nodata \\
\cross & 1.695 & \nodata & 5.5 & 2000-09-06 & 30 & 2001-12-08 & 10 & \nodata & \nodata \\

\enddata

\tablenotetext{a}{The Galactic \nh\ is based on \cite{d90}.}

\end{deluxetable}
\clearpage

\begin{deluxetable}{llclccclc}
\tabletypesize{\scriptsize}
\tablecolumns{9}
\tablewidth{0pt}
\rotate
\tablecaption{Fits to the spectra
of Gravitational Lensed Radio-quiet Quasars\label{tab:spec}}

\tablehead{
\colhead{} &
\colhead{} &
\colhead{} &
\colhead{} &
\multicolumn{3}{c}{Observed Properties\tablenotemark{b}} &
\colhead{} &
\colhead{} 
\\
\cline{5-7}
\\
\colhead{} &
\colhead{} &
\colhead{} &
\colhead{} & 
\colhead{$L_{X}$ (0.2--2 keV)} &
\colhead{$L_{X}$ (2--10 keV)} &
\colhead{$f_{X}$ (0.4--8 keV observed)} & 
\colhead{} & 
\colhead{}
\\
\colhead{Quasars} &
\colhead{Redshift} &
\colhead{Model\tablenotemark{a}} &
\colhead{$\Gamma$} &
\colhead{($\rm{erg~s^{-1}}$)} &
\colhead{($\rm{erg~s^{-1}}$)} &
\colhead{($\rm{erg~cm^{-2}~s^{-1}}$)} &
\colhead{$\chi^{2}_{\nu}(\nu)$} &
\colhead{$P(\chi^{2}/{\nu})$\tablenotemark{c}}
}

\startdata
HE 0230-2130 & 2.162 & pow & $1.93^{+0.08}_{-0.05}$ & $5.7\times10^{45}$ & $4.5\times10^{45}$ & $2.3\times10^{-13}$ & 1.09(39) & 0.33 \\
HS 0818+1227 & 3.115 & pow & $1.59^{+0.10}_{-0.08}$ & $2.9\times10^{45}$ & $4.4\times10^{45}$ & $1.3\times10^{-13}$ & 0.90(22) & 0.60 \\
APM 08279+5255\tablenotemark{d} & 3.911 & absori*(pow+zgau+zgau) & $1.78^{+0.04}_{-0.04}$ & $3.5\times10^{46}$ & $3.7\times10^{46}$ & $4.7\times10^{-13}$ & 0.86(235) & 0.95 \\
APM 08279+5255\tablenotemark{e} & 3.911 & absori*(pow+zgau+zgau) & $1.94^{+0.02}_{-0.02}$ & $5.3\times10^{46}$ & $4.3\times10^{46}$ & $5.0\times10^{-13}$ & 1.12(168) & 0.15 \\
RX J0911+0551 & 2.8  & zwabs*(pow+zgau) & $1.31^{+0.09}_{-0.10}$ & $9.0\times10^{44}$ & $2.6\times10^{45}$ & $1.1\times10^{-13}$ & 0.99(26) & 0.47 \\
LBQS 1009-0252 & 2.74 & pow & $1.9^{+0.2}_{-0.1}$ & $4.5\times10^{45}$ & $3.6\times10^{45}$ & $1.0\times10^{-13}$ & 1.57(9) & 0.12 \\
HE 1104-1805 & 2.303 & pow & $1.86^{+0.06}_{-0.04}$ & $4.4\times10^{45}$ & $4.0\times10^{45}$ & $1.8\times10^{-13}$ & 0.97(82) & 0.55 \\
PG 1115+080 & 1.72 & absori*(pow+zgau+zgau) & $1.90^{+0.03}_{-0.03}$ & $8.1\times10^{45}$ & $6.5\times10^{45}$ & $5.3\times10^{-13}$ & 1.15(200) & 0.07 \\
Q 1208+101 & 3.8 & pow+zgau & $2.3^{+0.2}_{-0.2}$ & $1.4\times10^{46}$ & $6.8\times10^{45}$ & $7.1\times10^{-14}$ & 1.13(6) & 0.34 \\
H 1413+117   & 2.55  & zwab*(pow+zgau) & $1.8^{+0.3}_{-0.3}$ & $4.0\times10^{45}$ & $4.5\times10^{45}$ & $1.0\times10^{-13}$ & 0.86(19) & 0.63 \\
HE 2149-2745\tablenotemark{f} & 2.033 & zwabs*pow & $1.4^{+0.6}_{-0.3}$ & $1.1\times10^{44}$ & $2.6\times10^{44}$ & $2.0\times10^{-14}$ & \nodata  & \nodata \\
Q 2237+0305 & 1.695 & zwabs*(pow+zgau) & $1.92^{+0.07}_{-0.06}$ & $6.5\times10^{45}$ & $5.9\times10^{45}$ & $4.6\times10^{-13}$ & 1.13(153) & 0.14 \\


\enddata

\tablecomments{The spectral fits were performed within the observed energy ranges 0.4--8 keV, and all derived errors are at the 68\% confidence level.}
\tablenotetext{a} {All models contain Galactic absorption model wabs, and all \chandra\ spectra contain ACISABS model to correct the contamination from the ACIS filter.  The pow, absori, zgau, and zwabs models represent powerlaw, ionized absorber, redshifted Gaussian emission or absorption feature, and redshifted absorption models, respectively.}
\tablenotetext{b} {The luminosity bandpasses are given in the rest-frame of the quasar.  The various absorption components are removed when calculating luminosites.  The observed values have not been corrected for the lensing magnification.}
\tablenotetext{c} {$P(\chi^{2}/{\nu})$ is the probability of exceeding $\chi^{2}$ for ${\nu}$ degrees of freedom.}
\tablenotetext{d} {The fit results correspond to the \chandra\ spectrum of \apm.}
\tablenotetext{e} {The fit results correspond to the \xmm\ spectrum of \apm.}
\tablenotetext{f} {The C statistic is used when fitting this low S/N spectrum. Other \chandra\ and \xmm\ spectra are binned with 15 and 100 events per bin, respectively.}

\end{deluxetable}
\clearpage

\begin{deluxetable}{llllll}
\tabletypesize{\scriptsize}
\tablecolumns{6}
\tablewidth{0pt}
\tablecaption{\aox\ Values of Gravitationally Lensed Quasars in the Sample\label{tab:aox}}

\tablehead{
\colhead{} &
\colhead{Optical} &
\colhead{} &
\colhead{} &
\colhead{} &
\colhead{} 
\\
\colhead{Quasars} &
\colhead{Information\tablenotemark{a}} &
\colhead{$f_{2500\AA}$\tablenotemark{b}} &
\colhead{$l_{2500\AA}$\tablenotemark{c}} &
\colhead{$f_{\rm 2keV}$\tablenotemark{d}} &
\colhead{\aox} 
}

\startdata
HE 0230-2130 & CASTLES & $0.37\pm0.07$ & 1.2 & $11.\pm1.$ & $-1.35\pm0.04$ \\
HS 0818+1227 & CASTLES & $1.9\pm0.4$ & 8.4 & $9.\pm1.$ & $-1.65\pm0.05$ \\
APM 08279+5255 & E00\tablenotemark{e} & $2.59\pm0.07$ & 13. & $5.4\pm0.4$ & $-1.80\pm0.01$ \\
RX J0911+0551 & CASTLES & $1.2\pm0.2$ & 4.9 & $2.3\pm0.3$ & $-1.81\pm0.04$ \\
LBQS 1009-0252 & CASTLES & $3.7\pm0.7$ & 15. & $32.\pm6.$ & $-1.56\pm0.05$ \\
HE 1104-1805 & CASTLES & $3.0\pm0.6$ & 10. & $11.0\pm0.6$ & $-1.70\pm0.03$ \\
PG 1115+080 & C87\tablenotemark{f} & $1.7\pm0.1$ & 4.4 & $11.9\pm0.4$ & $-1.59\pm0.01$ \\
Q 1208+101 & CASTLES & $11.\pm2.$ & 54. & $60.\pm20.$ & $-1.62\pm0.06$ \\
H 1413+117   & CASTLES & $1.6\pm0.3$ & 6.1 & $5.\pm3.$ & $-1.7\pm0.1$ \\
HE 2149-2745 & CASTLES & $7.\pm1.$ & 22. & $2.\pm1.$ & $-2.1\pm0.1$ \\
Q 2237+0305 & OGLE & $8.0\pm1.$ & 22. & $16.\pm1.$ & $-1.81\pm0.02$ \\

\enddata
\tablenotetext{a}{The programs or papers where we obtained the optical magnitudes used in the \aox\ calculation.}
\tablenotetext{b}{Rest-frame 2500~\AA\ flux density, in units of $10^{-27}~{\rm ergs~cm^{-2}~s^{-1}~Hz^{-1}}$.}
\tablenotetext{c}{Rest-frame 2500~\AA\ luminosity density, in units of $10^{30}~{\rm ergs~s^{-1}~Hz^{-1}}$.}
\tablenotetext{d}{Rest-frame 2~keV flux density, in units of $10^{-32}~{\rm ergs~cm^{-2}~s^{-1}~Hz^{-1}}$.}
\tablenotetext{e}{\citet{e00}.}
\tablenotetext{f}{\citet{c87}.}
\end{deluxetable}

\clearpage

\begin{deluxetable}{lcclll}
\tabletypesize{\scriptsize}
\tablecolumns{6}
\tablewidth{0pt}
\tablecaption{Magnification of Gravitationally Lensed Quasars in the Sample\label{tab:mag}}

\tablehead{
\colhead{} & 
\colhead{} & 
\colhead{} & 
\colhead{} & 
\colhead{$L_{X}$ (0.2 -- 2 keV)} &
\colhead{$L_{X}$ (2 -- 10 keV)}
\\
\colhead{Quasars} &
\colhead{Literature} &
\colhead{Model} &
\colhead{Flux Magnification} &
\colhead{($10^{44}~\rm{erg~s^{-1}}$)} &
\colhead{($10^{44}~\rm{erg~s^{-1}}$)} 
}

\startdata
HE 0230-2130 & B02\tablenotemark{d} & \nodata & 15 & 3.8 & 3.0 \\
HS 0818+1227 & B02 & \nodata & 10 & 2.9 & 4.4 \\
APM 08279+5255 & \nodata & SIE & 142 & 2.4 & 2.6 \\
RX J0911+0551 & \nodata & SIE & 17 & 0.53 & 1.5 \\
LBQS 1009-0252 & \nodata & SIE & 3.5 & 12.9 & 10.4 \\
HE 1104-1805 & \nodata & SIE & 12 & 3.6 & 3.3 \\
PG 1115+080 & T02\tablenotemark{b} & JF+CUSP+SIS\tablenotemark{c} & 26 & 3.1 & 2.5 \\
Q 1208+101 & B02 & \nodata & 3.1 & 45.8 & 21.7 \\
H 1413+117   & C99\tablenotemark{e} & \nodata & 23 & 1.7 & 2.0 \\
HE 2149-2745 & \nodata & SIE & 3.4 & 0.33 & 0.78  \\
Q 2237+0305 & S98\tablenotemark{a} & \nodata & 16 & 4.0 & 3.7 \\

\enddata

\tablenotetext{a}{\cite{s98}}
\tablenotetext{b}{\cite{t02}}
\tablenotetext{c}{JF, CUSP, and SIS are Jaffe, Cuspy, and Singular Isothermal Sphere mass models for the luminous mass of the galaxy, dark part of the galaxy, and nearby group of the system, respectively.}
\tablenotetext{d}{\cite{b02}}
\tablenotetext{e}{\cite{c99}}

\end{deluxetable}

\clearpage

\begin{deluxetable}{lccc}
\tabletypesize{\scriptsize}
\tablecolumns{6}
\tablewidth{0pt}
\tablecaption{Excess Variance of Gravitational Lensed Radio-quiet Quasars\label{tab:evar}}

\tablehead{
\colhead{} &
\colhead{$\sigma^{2}_{\rm{rms}}$} &
\colhead{$\sigma^{2}_{\rm{rms}}$} &
\colhead{$\sigma^{2}_{\rm{rms}}$} 
\\
\colhead{} &
\colhead{($10^{-2}$)} &
\colhead{($10^{-2}$)} &
\colhead{($10^{-2}$)}
\\
\colhead{Quasars} &
\colhead{Bin Size = 500 s} &
\colhead{Bin Size = 1000 s} &
\colhead{Bin Size = 2000 s}
}

\startdata
HE 0230-2130 & $-1.7\pm0.7$ & $-0.5\pm0.5$ & $-0.4\pm0.3$ \\
HS 0818+1227 & $0.1\pm2.2$ & $-0.6\pm2.0$ & $-0.3\pm0.8$ \\
APM 08279+5255 & $-0.5\pm0.3$ & $-0.3\pm0.2$ & $-0.1\pm0.1$\\
RX J0911+0551 & $-3.1\pm2.3$ & $-0.7\pm2.3$ & $0.9\pm2.9$ \\
LBQS 1009-0252 & $-0.5\pm2.6$ & $-3.1\pm0.8$ & $-1.4\pm0.9$ \\
HE 1104-1805 & $1.9\pm1.1$ & $1.1\pm0.8$ & $1.0\pm0.8$ \\
PG 1115+080 & $-0.2\pm0.5$ & $-0.2\pm0.3$ & $0.1\pm0.3$ \\
Q 1208+101 & $-6.1\pm1.5$ & $-2.7\pm0.9$ & $-1.2\pm0.8$ \\
H 1413+117   & $-2.4\pm3.5$ & $-3.0\pm2.5$ & $0.0\pm2.3$ \\
Q 2237+0305 & $-0.6\pm0.3$ & $-0.3\pm0.1$ & $0.0\pm0.2$ \\


\enddata

\end{deluxetable}

\clearpage

\begin{deluxetable}{lcccc}
\tabletypesize{\scriptsize}
\tablecolumns{5}
\tablewidth{0pt}
\tablecaption{Excess Variance of Quasars Observed with XMM\label{tab:evar2}}

\tablehead{
\colhead{} &
\colhead{$\sigma^{2}_{\rm{rms}}$} &
\colhead{$\sigma^{2}_{\rm{rms}}$} &
\colhead{$\sigma^{2}_{\rm{rms}}$} &
\colhead{$\sigma^{2}_{\rm{rms}}$} 
\\
\colhead{} &
\colhead{($10^{-2}$)} &
\colhead{($10^{-2}$)} &
\colhead{($10^{-2}$)} &
\colhead{($10^{-2}$)}
\\
\colhead{Quasars} &
\colhead{Bin Size = 500 s} &
\colhead{Bin Size = 1000 s} &
\colhead{Bin Size = 2000 s} &
\colhead{Bin Size = 3000 s}
}

\startdata
APM 08279+5255 & $0.4\pm0.2$ & $0.1\pm0.2$ & $0.1\pm0.2$ & $0.1\pm0.1$\\
RX J0911+0551 & $4.8\pm1.5$ & $3.2\pm1.5$ & $1.4\pm1.0$ & $-0.1\pm0.4$ \\
PG 1115+080 & $0.2\pm0.2$ & $0.1\pm0.2$ & $0.0\pm0.1$ & $-0.11\pm0.03$\\


\enddata

\end{deluxetable}
\clearpage

\clearpage

\begin{figure}
\epsscale{1}
\plotone{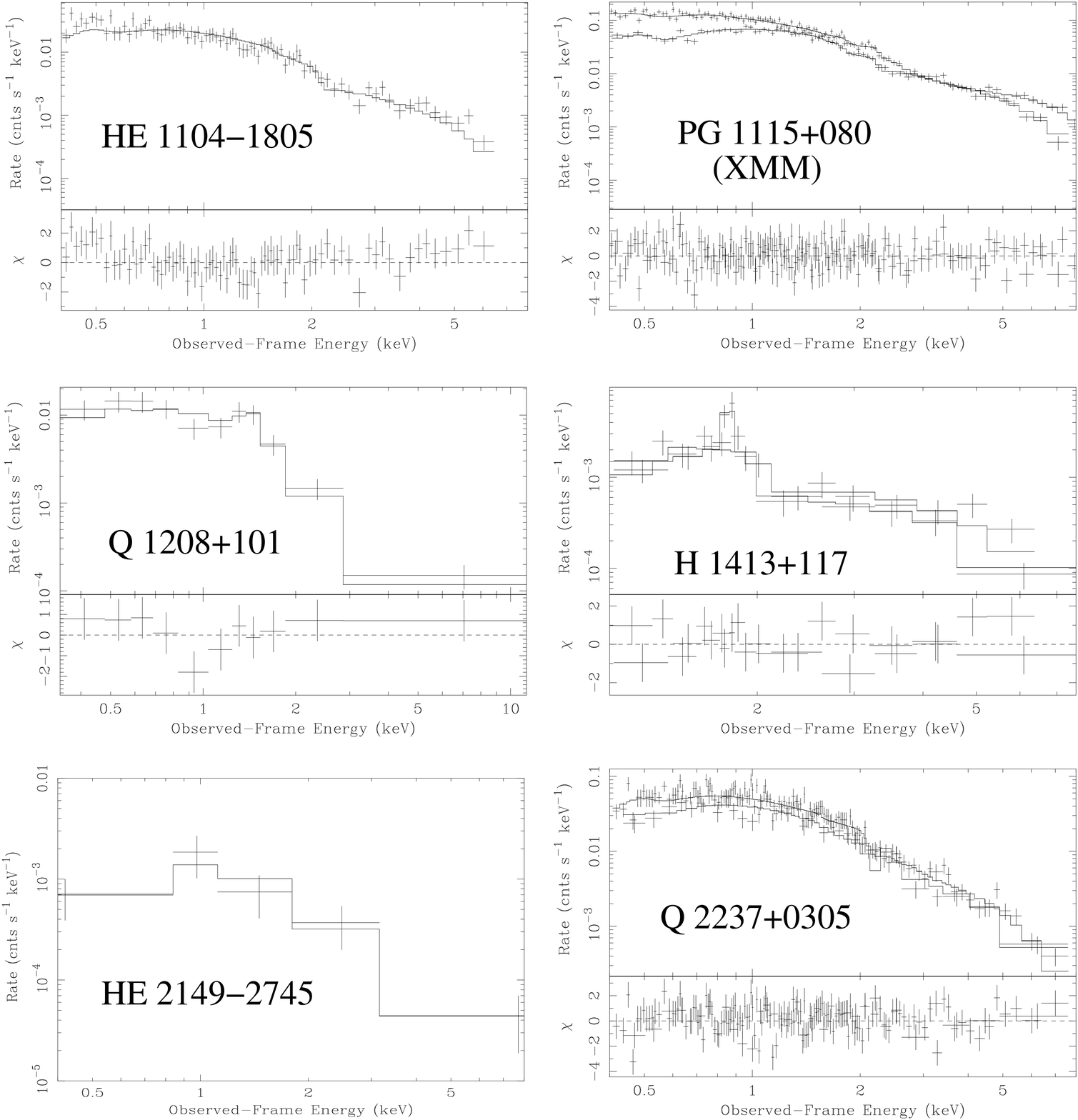}

Fig. 1. (continued)\label{fig:spec2}
\end{figure}
\clearpage

\begin{figure}
\epsscale{0.6}
\plotone{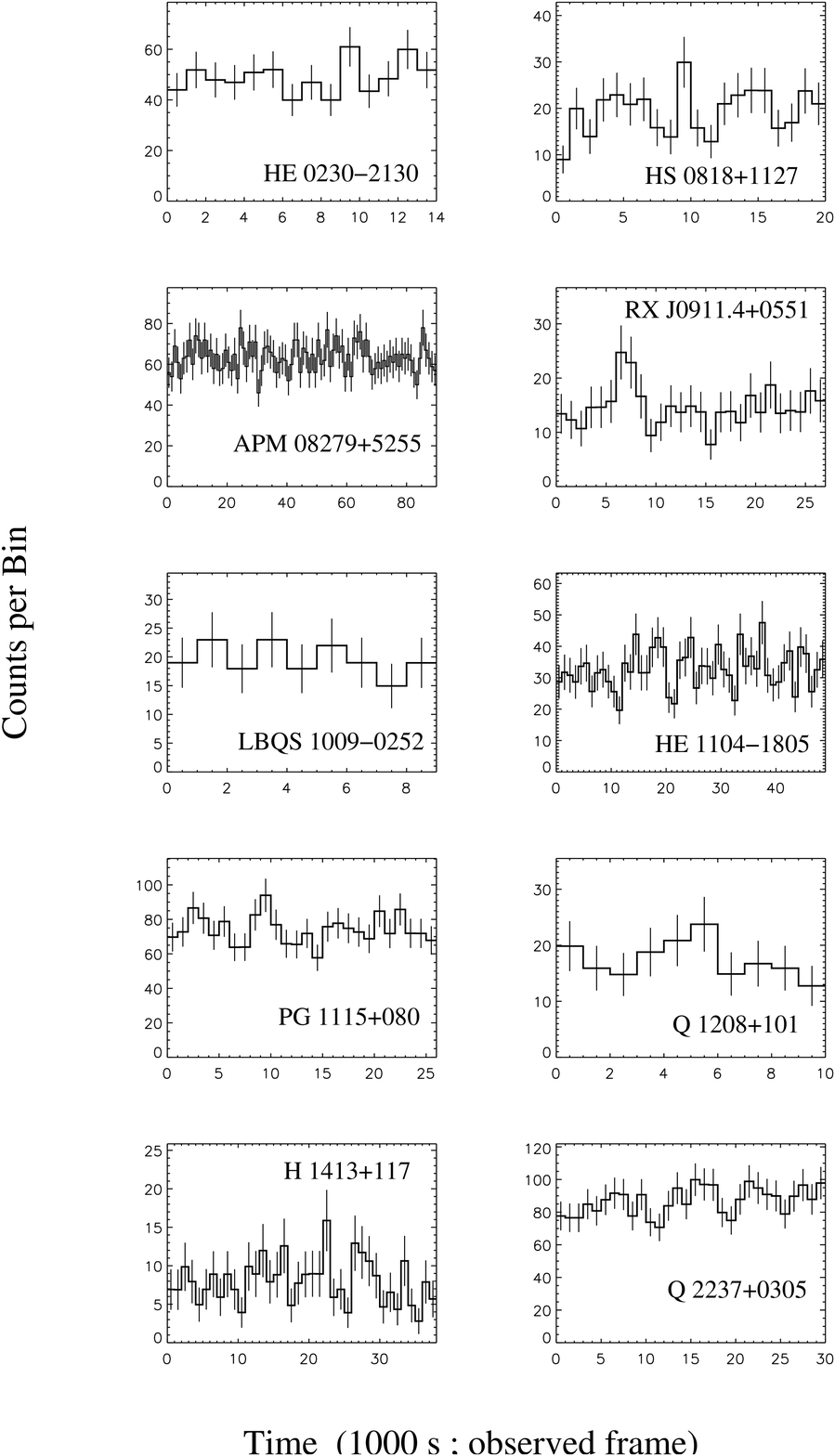}
\caption{Light-curves of gravitationally lensed radio-quiet quasars observed with \chandra.  The light-curves are binned with a bin size of 1000 s in the observed-frame.\label{fig:lc}}
\end{figure}
\clearpage

\begin{figure}
\epsscale{1}
\plotone{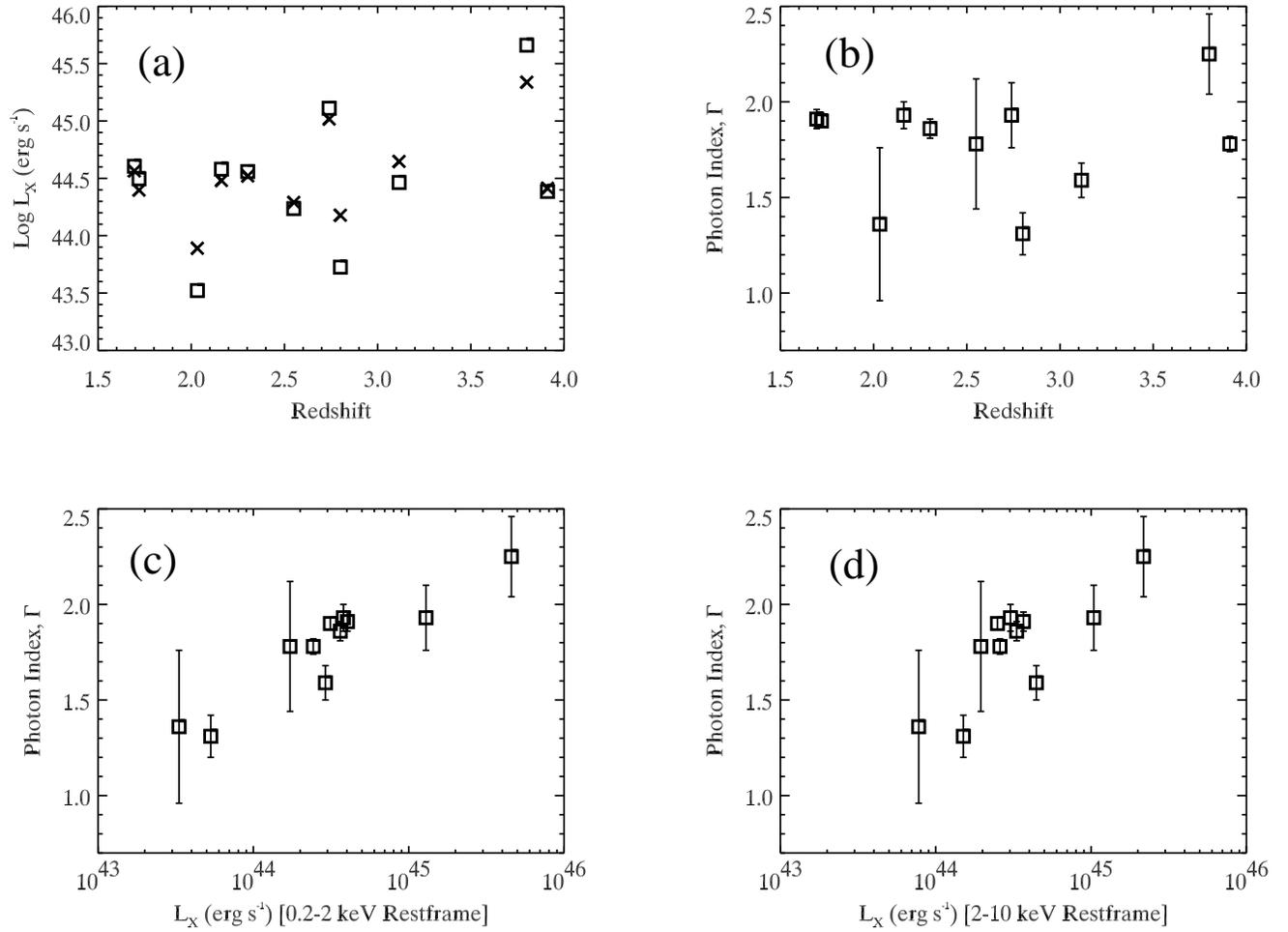}
\caption{Plots of (a) X-ray Luminosity vs. redshift.  The squares represent 0.2--2 keV luminosities and crosses represent 2--10 keV luminosities.  (b)X-ray photon index vs. redshift.  (c) X-ray photon index vs. 0.2--2 keV luminosity.  (d) X-ray photon index vs. 2--10 keV luminosity.\label{fig:gamma}}
\end{figure}
\clearpage

\begin{figure}
\epsscale{1}
\plotone{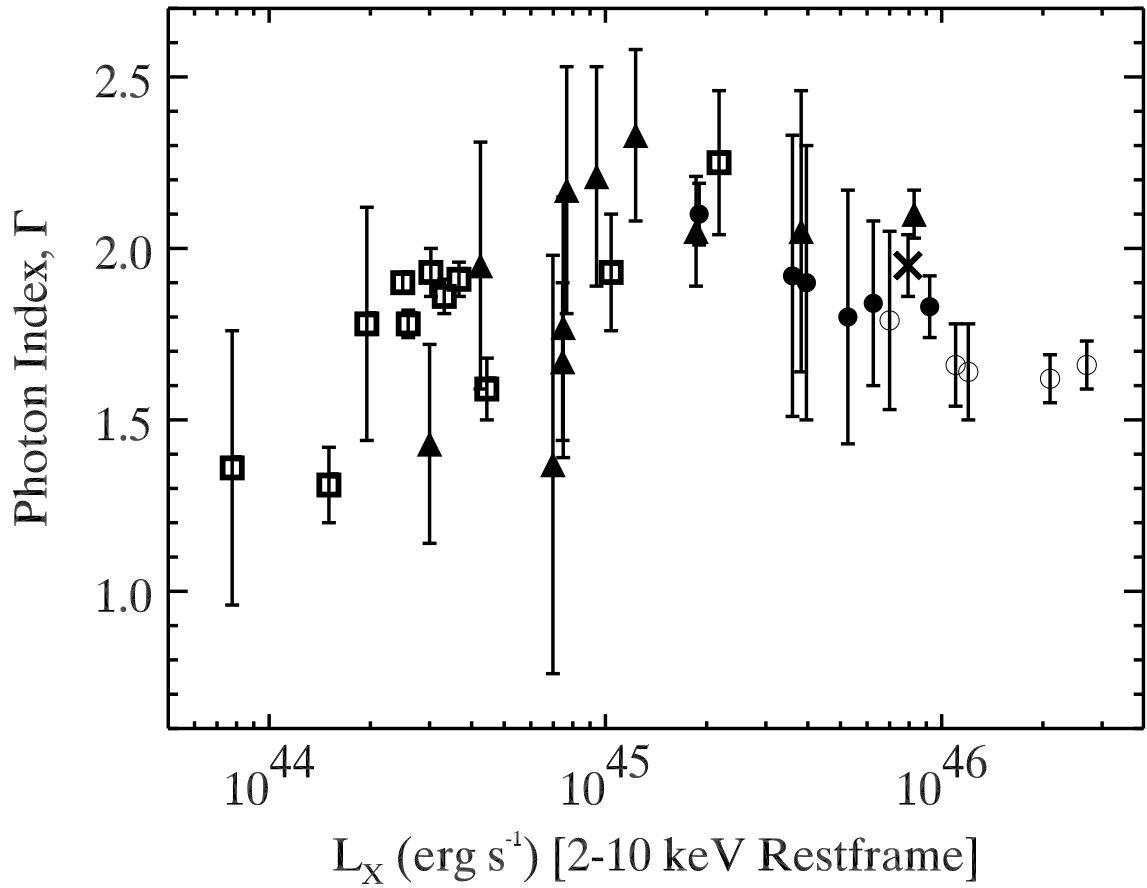}
\caption{$\Gamma$--$L_{X}$ (2--10 keV rest-frame) diagram for high redshift quasars ($z > 1.5$).  The data shown as filled triangles are from \citet{p03}.  The data shown as filled circles are from \citet{r00}.  The data shown as crosses are from \citet{g00}.  The data shown as open circles are from \citet{v99}.  The data shown as open squares are from our own sample of lensed quasars.\label{fig:gamma1}}
\end{figure}
\clearpage

\begin{figure}
\epsscale{0.8}
\plotone{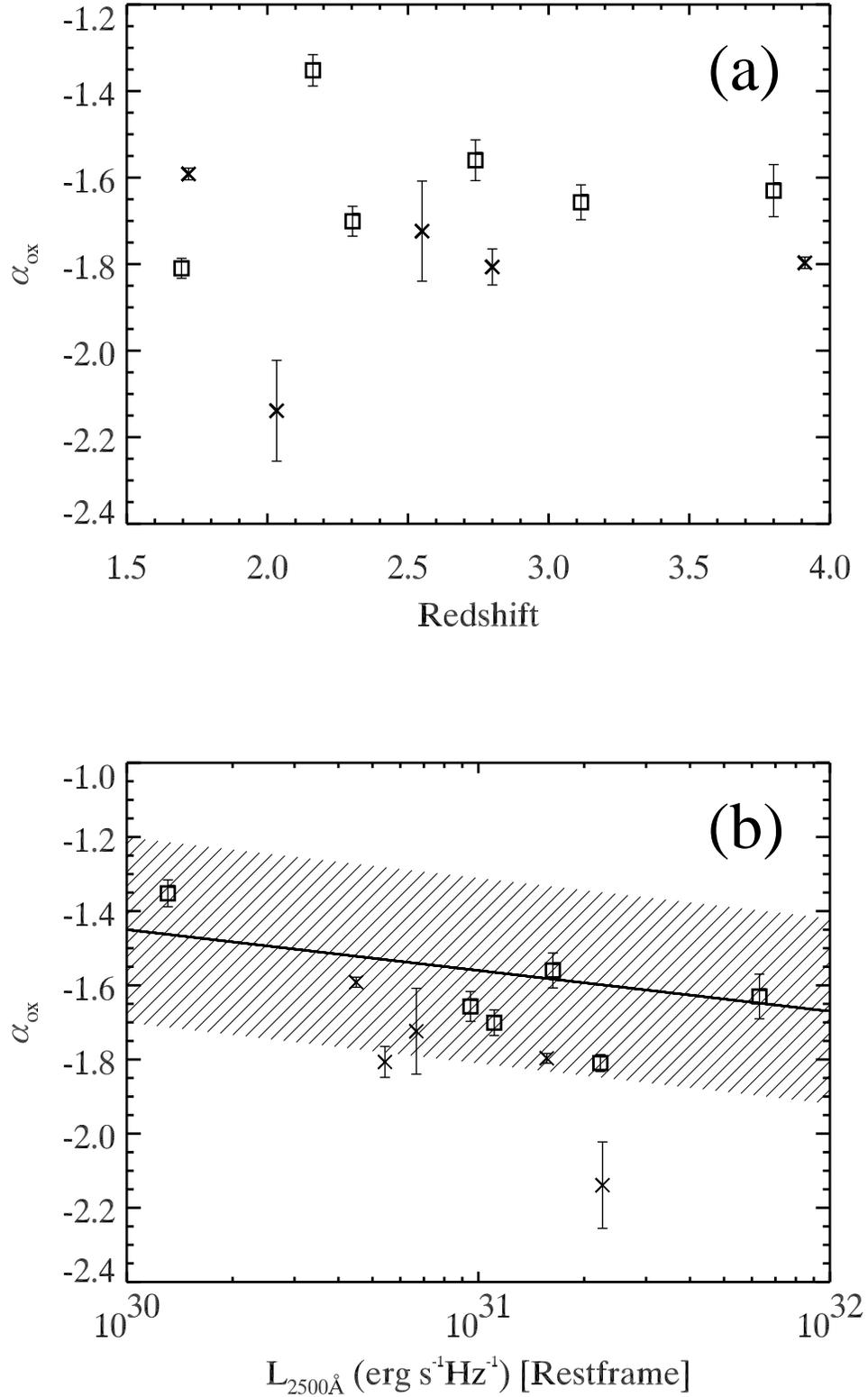}
\caption{Plots of \aox\ vs. redshift and UV luminosity.  The crosses represent BAL QSOs and squares represent non-BAL quasars for all of the plots in this figure. (a) \aox\ vs. redshift. (b) \aox\ vs. rest-frame 2500\AA\ luminosity.  The solid line represents the best fit \aox--$L_{2500\AA}$ relation obtained from \citet{vbs03}, and the shaded region represents a dispersion of 0.25 from the best fit line.\label{fig:aox}}
\end{figure}
\clearpage

\begin{figure}
\epsscale{1}
\plotone{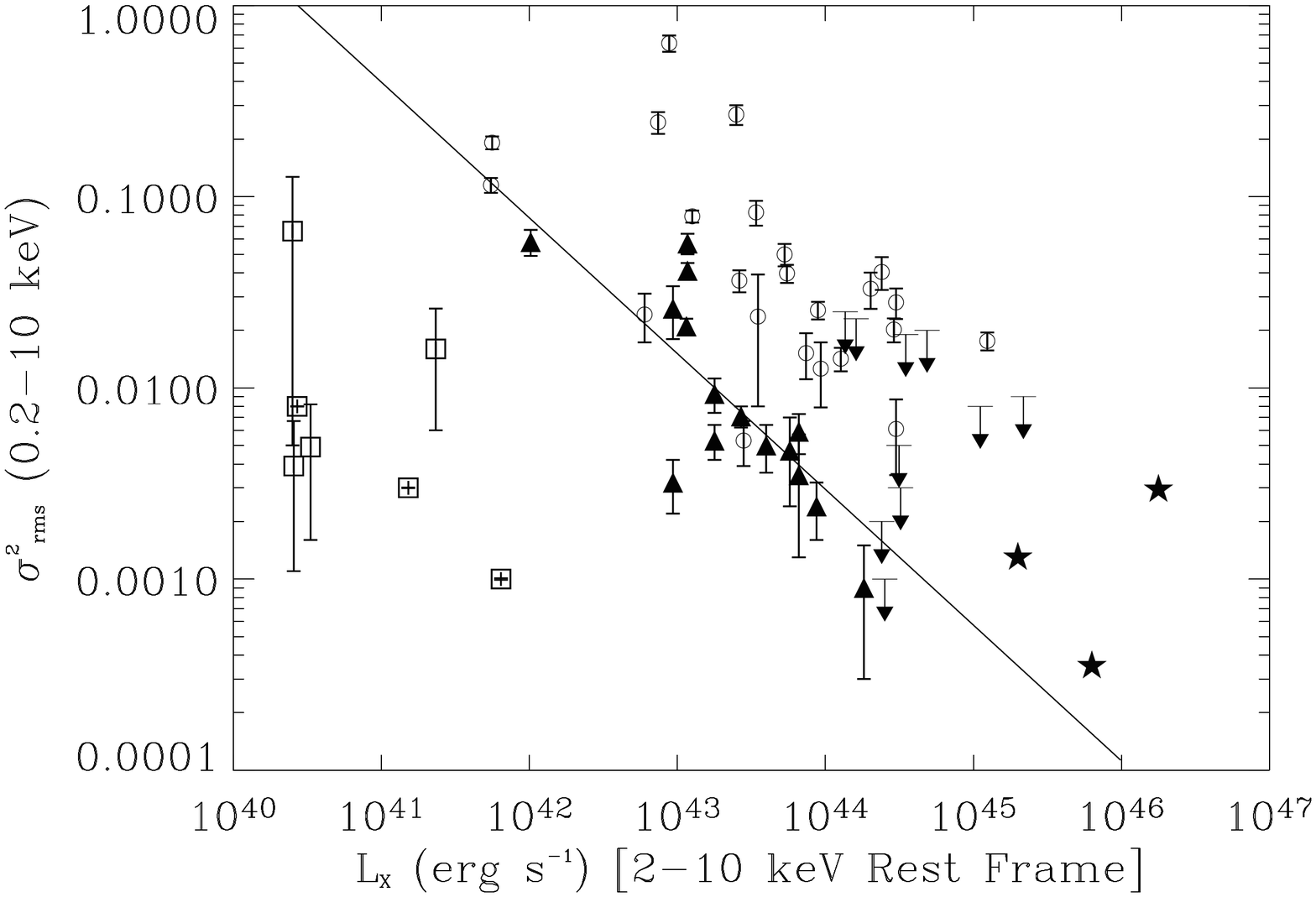}
\caption{Excess variance in the 0.2--10 keV band vs. 2--10 keV luminosity.   The data points representing the Seyfert 1s shown as filled triangles were obtained from \cite{n97} and are fitted with a power law (solid line).  The \asca\ SIS light-curves of the Seyfert 1s were binned in 128 s bins.  The data points for the NLS1s shown as open circles were obtained from \cite{l99}.  The \asca\ SIS light-curves of the NLS1s were binned in 128 s bins. The data points for the LINERS and LLAGNs shown as squares were obtained from \cite{p98}.  The data points shown as filled stars were obtained from $z > 2$ quasars of \cite{m02}. The downward arrows represent upper limits for the excess variance of the gravitationally lensed quasars from the present sample.\label{fig:excess}}
\end{figure}
\clearpage


\begin{thebibliography}{}

\bibitem[Agol, Jones, \& Blaes(2000)]{ag00} Agol, E., Jones, B., \& Blaes, O. 2000, \apj, 545, 657

\bibitem[Almaini \etal (2000)]{a00} Almaini, O., Lawrence, A., Shanks, T., Edge, A., Boyle, B. J., Georgantopoulos, I., Gunn, K. F., Stewart, G. C., \& Griffiths, R. E. 2000, \mnras, 315, 325

\bibitem[Arnaud(1996)]{a96} Arnaud, K. A. 1996, ASP Conf. Ser. 101: Astronomical Data Analysis Software and Systems V, ed. Jacoby G. \& Barnes J., 17

\bibitem[Barvainis, Alloin, \& Bremer (2002)]{b02a} Barvainis, R., Alloin, D., \& Bremer, M. 2002, \aap, 385, 399

\bibitem[Barvainis \& Ivison (2002)]{b02} Barvainis, R. \& Ivison, R. 2002, \apj, 571, 712

\bibitem[Bechtold \etal (2003)]{b03} Bechtold, J. \etal 2003, \apj, 588, 119

\bibitem[Boyle \etal (1987)]{b87} Boyle, B. J., Fong, R., Shanks, T., \& Peterson, B. A. 1987, \mnras, 227, 717

\bibitem[Brinkmann, Ferrero, \& Gliozzi (2002)]{bfg02} Brinkmann, W., Ferrero, E., \& Gliozzi, M. 2002, \aap, 385, L31

\bibitem[Chae \& Turnshek (1999)]{c99} Chae, K. \& Turnshek, D. A. 1999, \apj, 587, 597

\bibitem[Chartas, Brandt, \& Gallagher (2003)]{cbg03} Chartas G., Brandt, W. N., Gallagher S. C. 2003, \apj, in press, astro-ph/0306125 

\bibitem[Chartas \etal (2002)]{c02} Chartas, G., Brandt, W. N., Gallagher, S. C. \& Garmire G. P. 2002, \apj, 579, 169

\bibitem[Chartas \etal (2003)]{c03} Chartas, G., Brandt, W. N., Gallagher, S. C., \& Garmire G. P., 2003, submitted to ApJL

\bibitem[Chartas \etal (2001)]{c01} Chartas, G., Dai, X., Gallagher, S. C., Garmire, G. P., Bautz, M. W., Schechter, P. L., \& Morgan, N. D. 2001, \apj, 558, 119

\bibitem[Chiang \etal (2000)]{ch00} Chiang, J., Reynolds, C. S., Blaes, O. M., Nowak, M. A., Murray, N., Madejski, G., Marshall, H. L., \& Magdziarz, P. 2000, \apj, 528, 292

\bibitem[Christian, Crabtree, \& Waddell (1987)]{c87} Christian, C. A., Crabtree, D., \& Waddell, P. 1987, \apj, 312, 45

\bibitem[Dai \etal (2003)]{d03} Dai, X., Chartas, G., Agol, E., Bautz, M. W., \& Garmire, G. P. 2003, \apj, 589, 100

\bibitem[Dai \etal (2004)]{d04} Dai, X., \etal 2004, in preparation

\bibitem[Dickey \& Lockman (1990)]{d90} Dickey, J. M., \& Lockman F. J. 1990, ARA\&A 28, 215

\bibitem[Edelson \etal (2002)]{e02} Edelson, R., Turner, T. J., Pounds, K., Vaughan, S., Markowitz, A., Marshall, H., Dobbie, P., \& Warwick, R. 2002, \apj, 568, 610

\bibitem[Egami \etal (2000)]{e00} Egami, E., Neugebauer, G., Soifer, B. T., Matthews, K.,
 Ressler, M., Becklin, E. E., Murphy, T. W., Jr., \& Dale, D. A. 2000, \apj, 535, 561

\bibitem[Garmire \etal (2003)]{g03} Garmire, G. P., Bautz, M. W., Nousek, J. A., \& Ricker, G. R. 2003, SPIE, 4851,28

\bibitem[Georgantopoulos \& Papadakis (2001)]{g01} Georgantopoulos, I. \& Papadakis, I. E. 2001, \mnras, 322, 218

\bibitem[George \etal (2000)]{g00} George, I. M., Turner, T. J., Yaqoob, T., Netzer, H., Laor, A., Mushotzky, R. F., Nandra, K., \& Takahashi, T. 2000, \apj, 531, 52

\bibitem[Gliozzi, Sambruna, \& Eracleous (2003)]{gse03} Gliozzi, M., Sambruna, R. M., \& Eracleous, M. 2003, \apj, 583, 176

\bibitem[Haardt \& Maraschi (1993)]{h93} Haardt, F. \& Maraschi, L. 1993, \apj, 413, 507

\bibitem[Haardt, Maraschi, \& Ghisellini (1994)]{h94} Haardt, F., Maraschi, L., \& Ghisellini, G. 1994, \apj, 432, L95

\bibitem[Haardt, Maraschi, \& Ghisellini (1997)]{h97} Haardt, F., Maraschi, L., \& Ghisellini, G. 1997, \apj, 476, 620

\bibitem[ Holtzman \etal (1995)]{h95} Holtzman, J. A., Burrows, C. J., Casertano, S., Hester, J. J., Trauger, J. T., Watson,  A. M., \& Worthey, G. 1995, \pasp, 107, 1065

\bibitem[Impey \etal (1998)]{i98} Impey, C. D., Falco, E. E., Kochanek, C. S., Leh\'{a}r, J., McLeod, B. A., Rix, H.-W., Peng, C. Y., \& Keeton, C. R. 1998, \apj, 509, 551

\bibitem[Kauffmann \& Haehnelt (2000)]{kh00} Kauffmann, G, \& Haehnelt, M. 2000, \mnras, 311, 576

\bibitem[Keeton (2001a)]{k01a} Keeton, Charles R. 2001, astro-ph/0102340

\bibitem[Keeton (2001b)]{k01b} Keeton, Charles R. 2001, astro-ph/0102341

\bibitem[Laor \etal (1997)]{l97} Laor, A., Fiore, F., Elvis, M., Wilkes, B. J., \& McDowell, J. C. 1997, \apj, 477, 93

\bibitem[Leighly (1999)]{l99} Leighly, K. M. 1999, \apj, 125, S317

\bibitem[Manners, Almaini, \& Lawrence (2002)]{m02} Manners, J., Almaini, O., \& Lawrence, A. 2002, \mnras, 330, 390

\bibitem[Nandra \etal (1997)]{n97} Nandra, K., George, I. M., Mushotzky, R. F., Turner, T. J., \& Yaqoob, T. 1997, \apj, 476, 70

\bibitem[Page \etal (2003)]{p03} Page, K. L., Turner, M. J. L., Reeves, J. N., O'Brien, P. T., \& Sembay, S. 2003, \mnras, 338, 1004

\bibitem[Petrucci \etal (2000)]{p00} Petrucci, P. O., Haardt, F., Maraschi, L., Grandi, P., Matt, G., Nicastro, F., Piro, L., Perola, G. C., \& De Rosa, A 2000, \apj, 540, 131

\bibitem[Ptak \etal (1998)]{p98} Ptak, A., Yaqoob, T., Mushotzky, R., Serlemitsos, P., \& Griffiths, R. 1998, \apj, 501, L37

\bibitem[Reeves \& Turner (2000)]{r00} Reeves, J. N. \& Turner, M. J. L. 2000, \mnras, 316, 234

\bibitem[Reeves \etal (1997)]{r97}Reeves, J. N., Turner, M. J. L., Ohashi, T., \& Kii, T 1997, \mnras, 292, 468

\bibitem[Schlegel, Finkbeiner, \& Davis (1998)]{sch98} Schlegel, D. J., Finkbeiner, D. P., \& Davis, M. 1998, \apj, 500, 525

\bibitem[Schneider \etal (2001)]{sch01} Schneider, D. P., \etal 2001, \aj, 121, 1232

\bibitem[Schmidt, Webster, \& Lewis(1998)]{s98} Schmidt, R., Webster, R. L., \& Lewis, G. F. 1998, \mnras, 295, 488

\bibitem[Str\"{u}der \etal (2001)]{s01} Str\"{u}der, L., \etal 2001, \aap, 365, L18

\bibitem[Treu \& Koopmans (2002)]{t02} Treu, T. \& Koopmans, L. V. E. 2002, \mnras, 337, L6

\bibitem[Turner \etal (2001)]{t01} Turner, M. J. L, \etal 2001, \aap, 365, L27

\bibitem[Turner \etal (1999)]{t99} Turner, T. J., George, I. M., Nandra, K., \& Turcan, D. 1999, \apj, 524, 667

\bibitem[Vignali \etal (2003b)]{v03b} Vignali, C., Brandt, W. N., Schneider, D. P., Anderson, S. F., Fan, X., Gunn, J. E., Kaspi, S., Richards, G. T., \& Strauss, Michael A. 2003b, \aj, 125, 2876

\bibitem[Vignali, Brandt, \& Schneider (2003)]{vbs03} Vignali, C., Brandt, W. N., \& Schneider, D. P. 2003, \aj, 125, 433

\bibitem[Vignali \etal (2003a)]{v03a} Vignali, C., Brandt, W. N., Schneider, D. P., Garmire, G. P. \& Kaspi, S. 2003a, \aj, 125, 418

\bibitem[Vignali \etal (1999)]{v99} Vignali, C., Comastri, A., Cappi, M., Palumbo, G. G. C., Matsuoka, M., \& Kubo, H. 1999, \apj, 516, 582

\bibitem[Vanden Berk \etal (2001)]{v01} Vanden Berk, D. E., \etal 2001, \aj, 122, 549

\bibitem[Vaughan \& Edelson (2001)]{va01} Vaughan, Simon \& Edelson, Rick 2001, \apj, 548, 694

\bibitem[Yuan \etal (1998)]{y98} Yuan, W., Brinkmann, W., Siebert, J., \& Voges, W. 1998, \aap, 330, 108

\bibitem[Zdziarski \& Grandi (2001)]{z01} Zdziarski, Andrzej A. \&  Grandi, Paola 2001, \apj, 551, 186

\end{thebibliography}
\end{document}